\documentclass{mem}
\usepackage{natbib}\usepackage{txfonts}\usepackage{balance}
\usepackage{graphicx}
\usepackage[a4paper,breaklinks,dvipdfm]{hyperref}
\idline{75}{282}
\begin{document}

\title{The discovery of an anomalous RGB in M~2}

   \subtitle{}

\author{C.~Lardo
          \inst{1}, E.~Pancino\inst{2,3}, A.~Mucciarelli\inst{1}, and A.~P.~Milone\inst{4,5}
}

  \offprints{C. Lardo}

\institute{
Department of Physics and Astronomy --
University of Bologna, Via Ranzani 1,
I-40127 Bologna, Italy
\email{carmela.lardo2@unibo.it}
\and
Istituto Nazionale di Astrofisica --
Osservatorio Astronomico di Bologna, Via Ranzani 1, 
I-40127 Bologna, Italy
\and
ASI Science Data Center, I-00044 Frascati, Italy
\and
Instituto de Astrof\'{\i}sica de Canarias, 
E-38200 La Laguna, Tenerife, Canary Islands, Spain
\and 
Department of Astrophysics --
University of La Laguna, 
E-38200 La Laguna, Tenerife, Canary Islands, Spain
}

\authorrunning{C. Lardo}

\titlerunning{Discovery of an anomalous RGB in M~2}

\abstract{
Using $UV$ images taken with the Telescopio Nazionale Galileo, we discovered an anomalous sequence in the color-magnitude diagram of M~2. This feature appears as a narrow poor-populated red giant branch, which extends down to the sub giant branch region. We speculate that this new feature could be the extension of the faint component of the split sub giant branch recently discovered by Piotto et al.
We identified in our $U,V$ images two CH stars detected in previous studies.
These stars, which are both cluster members, fall on this redder sequence, 
suggesting indeed that the anomalous RGB should have a peculiar chemical pattern. 
Unfortunately, no additional spectra were obtained for stars in this previously unknown substructure.
}
\maketitle{}

\section{Introduction}\label{introduzione}
In the last decade, a large collection of new spectroscopic and photometric data had conclusively determined that Globular Clusters (GCs) are neither coeval nor monometallic, reopening the issue of the formation of such systems. The formation of GCs is now schematized as a two-step process, during which the enriched matter from the more massive stars -- both  intermediate mass asymptotic giant branch stars \citep{ventura08} or fast rotating massive stars \citep{decressin07} -- from a first generation gives birth, in the cluster innermost regions, to a  subset of stars with the characteristic signature of fully CNO-processed material \citep[see][and references therein]{gratton12}.
Variations in the light element abundances have been associated with the multiple sequences observed in the color-magnitude diagram (CMD), demonstrating that these  two phenomena are intimately linked  (e.g., \citealp{marino08,yongN}).
It is not surprising that the abundance variations should have large effects on photometry, particularly in the $UV$-blue bands where CN, NH, and CH molecular bands can be dominant. 
Here, we exploit the property of $U$-band observation of tracing light-element spreads and thus revealing multiple populations in the case of the poorly studied cluster M~2.

\section{Observations and data reduction}
We obtained images of the cluster in the standard Johnson 
$U$ and $V$ filters for a total of 540 s shifted in 3 single 
exposures in each filter with the DOLORES camera\footnote{DOLORES is a low resolution spectrograph 
and camera installed at Telescopio Nazionale Galileo (TNG) located in La Palma,
Canary Islands.}.
The choice of pass-bands is due to the ability of separating photometric sequences 
at different evolutionary stages along the CMD (see Sects.~\ref{introduzione} and~\ref{rgb_anomalo}).
The raw frames were processed using the
standard tasks in IRAF.
Point spread function (PSF) fitting photometry was hence carried out with the 
DAOPHOT II and ALLSTAR packages.
The photometric calibration was done using stars in common with Stetson Photometric standard 
field\footnote{{\tiny\tt http://www3.cadc-ccda.hia-iha.nrc-cnrc.gc.ca/\\community/STETSON/standards/}}.
Stars within 1\arcmin~and outside of 4\arcmin~from the cluster center are excluded from the CMD
to reduce blending effects and the field star contamination, respectively.
To select a sample of well-measured stars we have followed the procedure given in \citet{lardo12}, Sect.~5.1\footnote{We imposed also the selection limits of CHI $<$ 2.0 and -1 $<$ SHARP $<$ 1 photometric parameters.}.

\section{The anomalous RGB  in M~2}\label{rgb_anomalo}
Besides the remarkable exception of $\omega$ Centauri, \citep[see][and references therein]{jhonson10}, variations in the heavy element content have been detected only for few clusters.
NGC~1851 and M~22 are among the best studied of all these.
For these clusters, a bimodal distribution of $s$-process elements abundance have been identified
\citep{yong08, marino12}. The chemical inhomogeneity reflects itself in a complex CMD: multiple stellar groups in M~22 and NGC~1851 are also clearly manifest by a split in the sub giant branch (SGB) region \citep{piotto09,milone08} which appears to be related to chemical variations observed among red giant branch (RGB) stars \citep{marino12,lardo12}. Indeed, carefully constructed CMDs ---based on colors which include a blue filter \citep[][]{han09,marino12}---
clearly reveal that the bright SGB is connected to the blue RGB, while red RGB stars are linked to the faint SGB.
The split of the RGB discovered in the $U-I$ and $U-V$ colors for NGC~1851 and M~22 respectively,
would not be detected in usual optical colors.

\begin{figure}
  \centering
  \resizebox{\hsize}{!}{\includegraphics{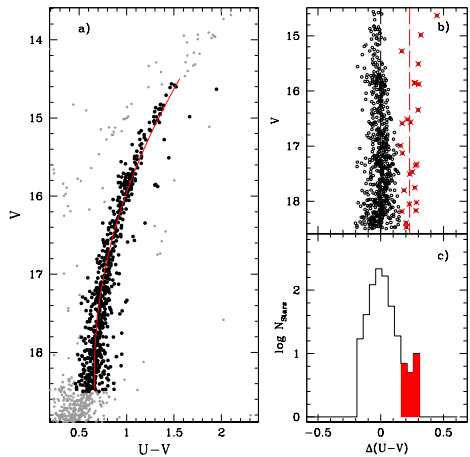}}
\caption{\footnotesize{(a) $U, V$ CMD from DOLORES images is shown in gray. Selected RGB stars are plotted as black dots, while the red continuous line
is the fiducial obtained in the way described in the text. Panels (b) and (c) show the rectified RGB in function of the color difference and the histogram color distribution respectively (see the text for details).}}
        \label{DOLO}
   \end{figure}

M~2 DOLORES photometry (see Fig.~\ref{DOLO}) displays an {\em anomalous} branch beyond the red edge of the main body of the RGB. The difference in color between stars belonging to this structure and {\em normal} RGB stars 
is quite large (of the order of 0.2-0.3 mags, well above the typical measurement errors) and extends down to the SGB region.
There may be a second group of stars that are 0.3 mags redder with respect to this sequence and can possibly be more, anomalous RGB stars.
 \begin{figure}
  \centering
  \resizebox{\hsize}{!}{\includegraphics{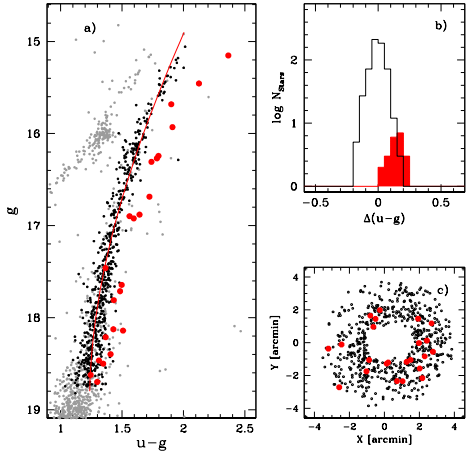}}
\caption{\footnotesize({a) $u, (u-g)$ CMD from \citet{an08} {\em corrected} photometry zoomed in around the RGB. Stars selected as red in Fig.~\ref{DOLO} are  plotted as red circles, while the red continuous line is the fiducial obtained in the way described in the text. Panels (b) 
show the color distribution in the $(u-g)$ color, while panel (c) show spatial distribution of the selected red stars.}}
        \label{CHECK}
   \end{figure}
We eliminated unphysical effects (i.e., differential reddening and field contamination) as the cause of this observed additional RGB (for details refer to \citealp{lardoM2}).
To take photometric errors into due account, we follow the method described in \citet{anderson09}.
We considered the two independent CMDs obtained from DOLORES and \citet{an08} photometry.
In Fig.~\ref{DOLO} we selected the portion of the RGB sequence with magnitudes between $14.5 \leq V \leq 18.5$ mag. In addition, we defined {\em bona fide} RGB members as the stars closer to the main RGB locus in the corrected DOLORES CMD (showed as black dots in panel (a) of Fig~\ref{DOLO}). The red continuous line in the same panel is the RGB fiducial obtained following as described in \citet{milone08}.
In brief, we drew a ridge line by putting a best-fit spline through the average color computed
in successive short (0.2 mag) magnitude intervals. In panel (b) we subtracted from the color of each star the color of the fiducial at the same magnitude and plotted the $V$ magnitude in function of this color difference; $\Delta (U-V)$.
The histogram color distribution in logarithmic scale in panel (c) presents a clear substructure at the red end of the RGB, and we arbitrarily isolated RGB stars with $\Delta (U-V)>$0.15 (red shaded region). These stars are plotted as red crosses in panel (b).

If the red branch we see was due to photometric errors, then a star redder than the RGB ridge line in the $V, (U-V)$ diagram would have the same probability of being bluer or redder in a different CMD obtained with different data. To this purpose, we identified the selected stars in {\em u,g} photometry \citep{an08} in Fig.~\ref{CHECK} (red dots). The (a) panel shows a zoom around the RGB, again the red line is the fiducial defined as discussed above.
In the following analysis, we considered only stars in common with the DOLORES photometry and, for the sake of homogeneity, we kept only stars between 1\arcmin $<R<$4\arcmin~from the cluster center.
The fact that the histogram distributions of the selected RGB stars systematically have red colors demonstrates that we are seeing a {\em real} feature: no random or systematic errors can explain why the two distribution remain confined in the CMDs obtained from independent data-sets.
Similar spatial distributions of stars on the bluer  and redder RGBs also indicate that the differential reddening, if any, is not likely the cause of the double RGBs (see panel (c) in Fig.~\ref{CHECK}).
 \begin{figure}
  \centering
  \resizebox{\hsize}{!}{\includegraphics{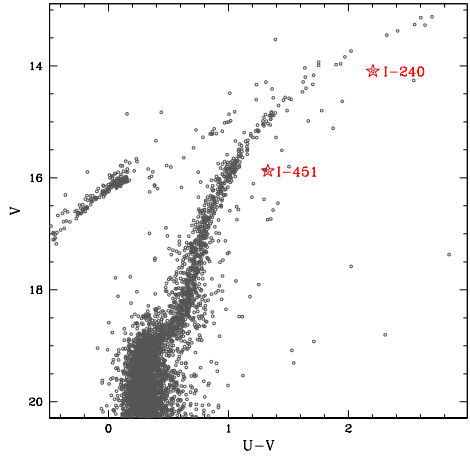}}
\caption{\footnotesize{CMD for M2. The location of the carbon stars in
the CMD is indicated by the open stars.}}
        \label{CHSTAR}
   \end{figure}

We found that the stars located on the anomalous red substructure 
account for only $\simeq$ 4\% of the whole RGB population in this range of magnitude ($14.5 \leq V \leq 18.5$ mag).
For comparison, $\simeq 5\%$ of stars turn out to belong to the faint SGB if the two subpopulations have the same age and  twice C+N+O and $\simeq 3\%$ if they have the same C+N+O but differ of $\simeq$ 1--2 Gyr in age \citep{piotto12}.
Moreover \citet{piotto12} claimed for this cluster 
the presence of a split SGB, with a fainter component less populous than the brighter one.
We tentatively speculate that, also for M~2, this newly discovered double RGB might be photometrically connected to the split SGB, in close analogy to the case of NGC~1851 and M~22. 
\subsection{CH stars along the anomalous RGB}\label{CH}
M~2 contains two CH stars, as discovered by \citet{zinn81} and \citet{smith90}.
These stars show abnormally high CH absorption, together with deep CN bands, compared to other cluster giants. 
They are seen in dSph galaxies, and in the Galactic halo, but they
are relatively rare within globular clusters.
In Fig.~\ref{CHSTAR} we identified the two CH stars 
in our $V, U-V$ photometry.
Interestingly enough, both stars belong to the additional RGB, pointing out the anomalous chemical nature
of this redder branch. Regardless of the exact classification of I-240 and I-451, it is apparent that the anomalous RGB contains a populations of giants which exhibit both a strong CN and strong G band. These stars may be the analogous to  other CN and CH-strong RGB stars found in $\omega$~Cen, M~22 and NGC~1851 \citep{hesser82}.
Given the peculiarity of other clusters that contain CH stars, it is of extreme interest to investigate the chemical pattern of stars in this red substructure. 
\section{Conclusions}\label{conclusioni}
Among the GCs with photometric evidence of multiple populations only NGC~1851 and M~22 display a bimodal SGB  which is photometrically connected to the split RGB 
      \citep{lardo12,marino12}.
        The apparent similarity of M~2 to NGC~1851 and M~22 calls for a deeper
      and complete spectroscopic characterization of stars in this anomalous RGB:
only accurate measurements of metal abundances for a representative sample of stars will shed light on the 
origin of this poorly studied cluster.
\begin{acknowledgements}
Support for this work has been provided by the IAC (grant 310394),
and the Education and Science Ministry of Spain (grants AYA2007-3E3506, and AYA2010-16717).
\end{acknowledgements}

\bibliographystyle{aa}
\bibliography{m2_msai.bib}

\end{document}